\documentclass[pra,twocolumn,superscriptaddress,showpacs]{revtex4-1}
\usepackage{amssymb,amsmath,graphicx,color}
\usepackage{graphicx,epsfig}
\usepackage{times}

\begin{document}
\title{Quantum state conversion in opto-electro-mechanical systems via shortcut to adiabaticity}

\author{Xiao Zhou}
\author{Bao-Jie Liu}
\affiliation{Guangdong Provincial Key Laboratory of Quantum Engineering and Quantum Materials, and School of Physics\\ and Telecommunication Engineering, South China Normal University, Guangzhou 510006, China}

\author{L.-B. Shao}
\affiliation{National Laboratory of Solid State Microstructures and School of Physics, Nanjing University, Nanjing 210093, China}

\author{Xin-Ding Zhang}\email{xdzhang@scnu.edu.cn}
\author{Zheng-Yuan Xue}\email{zyxue@scnu.edu.cn}
\affiliation{Guangdong Provincial Key Laboratory of Quantum Engineering and Quantum Materials, and School of Physics\\ and Telecommunication Engineering, South China Normal University, Guangzhou 510006, China}

\date{\today}

\begin{abstract}
Adiabatic process has found many important applications in modern physics, the distinct merit of which is that it does not need accurate control over the timing of the process. However, it is a slow process, which limits the application in quantum computation, due to the limited coherent times of typical quantum systems. Here, we propose a scheme to implement quantum state conversion in opto-electro-mechanical systems via shortcut to adiabaticity, where the process can be greatly speeded up while the precise timing control is still not necessary. In our scheme, only by modifying the coupling strength, we can achieve fast quantum state conversion with high fidelity, where the adiabatic condition does not need to be met. In addition, the population of the unwanted intermediate state can be further suppressed. Therefore, our protocol presents an important step towards practical state conversion between optical and microwave photons, and thus may find many important applications in hybrid quantum information processing.
\end{abstract}


\maketitle

\section{Introduction}
Hybrid quantum systems \cite{hybrid} may consolidate the advantages of different systems, and thus may find many important applications in quantum information processing. Recently, opto-electro-mechanical systems \cite{oem1,oem2} that interfacing optical and microwave photons have attracted considerable attention due to the advanced fabrication of superconducting circuits that support microwave photons and the scalable integrated optical photonic circuit techniques.
Through opto-electro-mechanical  systems, one can efficient up-conversion microwave information to the optical counterpart, and thus enable the transmission of the information through optical fibres in a low-loss way. Therefore, great efforts have been paid to the conversion between microwave and optical fields
\cite{Wang,tian,science,Hill,tianlin,Lukin,NJP,xue1,xue2,xue3,wang2}. However, the conversion process usually uses the adiabatic passage, which requires a long operation time to satisfy the adiabatic criteria, and thus decoherence may induce unacceptable loss.

One possible way out of the difficulty is the so-called "superadiabatic transitionless driving (SATD)" \cite{Berry,TQD1,TQD2} or "shortcut to adiabatic (STA)'' \cite{chenxi,PRL2012,PRA2016,oe2016,PRA2015} protocol, where the conversion process is speeded up and still keep the merits of the adiabatic passage. In this protocol, a system will force to follow exactly the instantaneous eigenstates of its Hamiltonian by applying additional a precisely controlled field to cancel nonadiabatic transitions between the instantaneous eigenstates  \cite{Chen2012,iterative,Chen2014,s1,s2,s3,song,s4,Deng}. In particular, it is found that the protocol can still be simplified \cite{du,Y.C.Li,dress,xiayan,15,zhangshou,16,liubaojie}, only by modifying the driving fields of the adiabatic case. Moreover, it is also indicated that the populations of  the unwanted intermediate state can be suppressed by properly choosing the control parameters, thus reduces its influence and leads to higher fidelity \cite{dress,17}.

Here, we propose a scheme to achieve quantum state conversion (QSC) between microwave and optical modes in opto-electro-mechanical systems via shortcut to adiabaticity. Our scheme holds the advantages of the QSC in an adiabatic way but does not require the adiabatic condition to be met, and thus has potential applications in hybrid quantum information processing.

\section{The system and its adiabatic dynamics}
We consider an opto-electro-mechanical system, as illustrated in Fig.1(a), where a mechanical resonator simultaneously coupled to an optical cavity
and a microwave cavity via dispersive coupling and each cavity modes are under external driving with frequencies $\omega_{d_i}$ ($i$=1, 2) \cite{optomechanics,optomechanics2}. Follow the standard linearization procedure \cite{optomechanics}, the system can be described by
\begin{align}\label{1}
 H=\omega_m\hat{b}_m^\dag\hat{b}_m + \sum_{i=1}^2 \left[ \Delta_i\hat{a}_i^\dag \hat{a}_i  + g_i (\hat{a}_i^\dagger\hat{b}_m+\hat{a}_i\hat{b}_m^\dagger)\right],
\end{align}
where we have assumed $\hbar =1$ (here and hereafter); $\hat{a}_1$, $\hat{a}_2$ and $\hat{b}_m$ are the annihilation operators for the optical, microwave and mechanical modes; $\omega_i$ ($i$=1, 2) is the frequency of the $ith$ cavity mode, $\omega_m$ is the mechanical oscillator's frequency, and $\Delta_i=\omega_{di}-\omega_i$ is the detuning between cavity mode and external driving; $g_i=G_{0i}\sqrt{n_i}$ ($i$=1, 2) is the effective linear coupling that is proportional to the driving amplitude applied to cavity $i$, which are tunable by varying the driving field \cite{tunable coupling} with $G_{0i}$ and $n_i$ are the effective single-photon coupling strength and photon number inside the cavity, respectively.  Note that the Hamiltonian in Eq. 1 is written in a displaced frame, which has a form readily for QSC,  and thus the quantum state to be transferred here sits atop a classical coherent state with large number of photons \cite{Wang}.

\begin{figure}[tbp]
\begin{center}
\includegraphics[width=8cm]{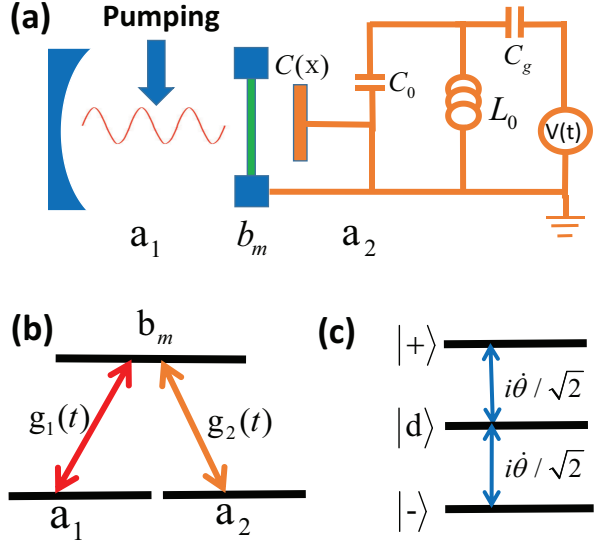}
\caption{Illustration of the proposed protocol. (a) Schematic diagram for the two cavity optomechanical system, the electromechanical coupling is induced by a capacitance $C(x)$ of the microwave cavity $a_2$,  and $C_g$ is the gate capacitance of the microwave cavity $a_2$. (b) Coupling configuration in the single excitation subspace of the considered system.   (c) Energy level in the dressed picture, where $\dot\theta(t)$ stand for the nonadiabatic transition between $|d\rangle$ and $|+\rangle$, $|-\rangle$.}
\end{center}
\end{figure}

We consider that the case of strong coupling with external driving in the first red sideband of the mechanical mode, i.e., $-\Delta_i=\omega_m$. Moving to the interaction picture, and using the rotating wave approximation, we obtain that
\begin{equation}\label{hint}
H_{int}=
 \left(
   \begin{array}{ccc}
    0 & g_1(t) & 0 \\
     g_1(t) & 0 &  g_2(t)\\
     0 & g_2(t) &  0 \\
   \end{array}
 \right),
\end{equation}
where we have assumed the basis vectors as
$|a_1\rangle=[1,0,0]^T$, $|b_m\rangle=[0,1,0]^T$, and $|a_2\rangle=[0,0,1]^T$, corresponding to states with one excitation of  the optical cavity, the mechanical oscillator and the microwave cavity, respectively. Noted that the state transfer protocol is confined in single excitation subspace formed by $\{|a_1\rangle, |b_m\rangle, |a_2\rangle\}$ in low temperatures. Three instantaneous eigenvectors of the Hamiltonian in Eq. (\ref{hint}) can be described by
\begin{align}
|+\rangle&=\frac{1}{\sqrt2}\left(\sin\theta|a_1\rangle
+|b_m\rangle+\cos\theta|a_2\rangle\right),\notag\\
|d\rangle&=-\cos\theta|a_1\rangle+\sin\theta|a_2\rangle,\notag\\
|-\rangle&=\frac{1}{\sqrt2}\left(\sin\theta|a_1\rangle-|b_m\rangle+\cos\theta|a_2\rangle\right),
\end{align}
and the corresponding three eigenvalues are $E_-=-g(t)$, $E_d=0$, and $E_+=g(t)$, where $g(t)=\sqrt{g_1^2(t)+g_2^2(t)}$ and $\tan\theta=g_1(t)/g_2(t)$, as shown in Fig.1(c). In this system, $|+\rangle$ and $|-\rangle$ are the "bright'' mode, which are superpositions of the cavity $|a_1\rangle$, cavity$|a_2\rangle$ and the mechanical mode $|b_m\rangle$; while $|d\rangle$  is the "dark'' mode, which decouples from the mechanical mode $|b_m\rangle$ due to destructive interference. If one initially prepared a excitation in the microwave mode $|a_2\rangle$, through the "dark'' state passage, one can adiabatically convert the excitation to the optical mode $|a_1\rangle$, and vice versa \cite{science}.

\section{Shortcut to adiabatic quantum state conversion}
\subsection{The protocol}
We now consider the case of speeding up the adiabatic QSC process, where the adiabatic condition is not met.  In the adiabatic basis $\{|+\rangle, |d\rangle, |-\rangle\}$, the Hamiltonian in Eq. (\ref{hint}) becomes
 \begin{align} \label{had}
 H_{ad}(t)&=U(t)H_{int}U^\dagger(t)+i\frac{dU(t)}{dt}U^\dagger(t) \notag\\
 &=g_0 \hat{M}_z-\dot{\theta} \hat{M}_y,
 \end{align}\\
where
\begin{equation}
    U(t)={1 \over \sqrt{2}}
  \left(
    \begin{array}{ccc}
      \sin\theta& 1 &  \cos\theta\\
      -\sqrt{2}\cos\theta & 0 & \sqrt{2}\sin\theta \\
     \sin\theta & -1  &\cos\theta \\
    \end{array}
  \right),
   \end{equation}
$\hat{M_x}=(|-\rangle-|+\rangle)\langle d|/\sqrt{2}+H.c.$, $\hat{M_y}=i(|+\rangle+|-\rangle)\langle d|/\sqrt{2}+H.c.$,
and $\hat{M_z}=(|+\rangle\langle+|-|-\rangle\langle-|)$ are the pauli matrix for spin 1 system, which obey the commutation relation $[\hat{M}_p,\hat{M}_q]=i \varepsilon^{pqr} \hat{M}_r$ with $\varepsilon^{pqr}$ being the Levi-Civita symbol.  The second term of the Hamiltonian in Eq. (\ref{had}) corresponds to nonadiabatic transitions among the dark state and the two bright states when the adiabatic condition $|\dot{\theta}|\ll g(t)$ is not met well, as shown in Fig. 1 (c).

In order to correct the nonadiabatic leakage, a correction Hamiltonian
\begin{align}\label{8}
 H_c^t =\left(
                          \begin{array}{ccc}
                            0 & 0 & i\dot{\theta}  \\
                            0 & 0 &0 \\
                           -i\dot{\theta} &0 &0 \\
                          \end{array}
                        \right).
\end{align}
is introduced, so that the total Hamiltonian becomes
\begin{equation}\label{7}
H_{m}^t=H_{int}+H_c^t,
\end{equation}
for transitionless quantum driving. Therefore, in the adiabatic basis, the total modified Hamiltonian becomes
 \begin{align}
  H_{m}^t &=U(t)\left[H_{ad}(t)+H_c^t(t)\right]U^\dagger(t)
  +i\frac{dU(t)}{dt}U^\dagger(t)\notag\\
  &=g(t)\hat{M}_z,
 \end{align}
which does not have the nonadiabatic transitions. However, the Hamiltonian in Eq. (\ref{8}) refers to the direct coupling between the microwave and optical modes, which is hard to be directly induced experimentally.

To overcome this obstacle, we look for another correction Hamiltonian $H^{d}_{c}$ via the dressed state method \cite{dress}, so that we can speed up the QSC by only modifying the pulse shape of the coupling strength in Eq. (\ref{hint}). Here, we take
\begin{align}
H_c^d=U^\dagger(t)\left[g_x(t)\hat{M}_x+g_z(t)\hat{M}_z\right]U(t),
\end{align}
so that the total Hamiltonian in the adiabatic basis now becomes
\begin{align}
 \tilde{H}(t)&=H_{int}+H_c^d\notag\\
 &=\tilde{g}_{1}(t)|b_m\rangle\langle a_1|+\tilde{g}_{2}(t)|b_m\rangle\langle a_2|+ \text{H.c.},
\end{align}
where the modified couplings are
\begin{align}
\tilde{g}_{1}(t)=[g(t)+g_z(t)]\sin\theta+g_x(t)\cos\theta,\notag\\
\tilde{g}_{2}(t)=[g(t)+g_z(t)]\cos\theta-g_x(t)\sin\theta.
\end{align}
From the discussion in the last section, we know that this can not be achieved if we use the adiabatic eigenstates as the conversion channel. However, it only requires the initial and  the final states are in the adiabatic eigenstates. Therefore, we move to the dressed state picture with respect to $\hat{V}(t)=\exp[i\mu(t)\hat{M_x}]$. After these two transformations, we find that the total Hamiltonian becomes
\begin{align}
H_{m}^d &=VH_{ad}(t)V^\dagger +VUH_c^d U^\dagger V^\dagger+i\frac{dV}{dt}V^\dagger.
\end{align}
As the modified Hamiltonian $H_m^d$ should be designed to cancel out the unwanted off-diagonal elements, the controlled parameters $g_x(t), g_z(t)$ should be
\begin{align}\label{15}
g_x(t)=\dot{\mu}(t),\quad g_z(t)=-g(t)+ \dot{\theta} /\tan\mu(t).
\end{align}
where, for a simplest nontrivial example, we may set $g_z(t)=0$.

\begin{figure}[tbp]
\begin{center}
    \includegraphics[width=8.5cm]{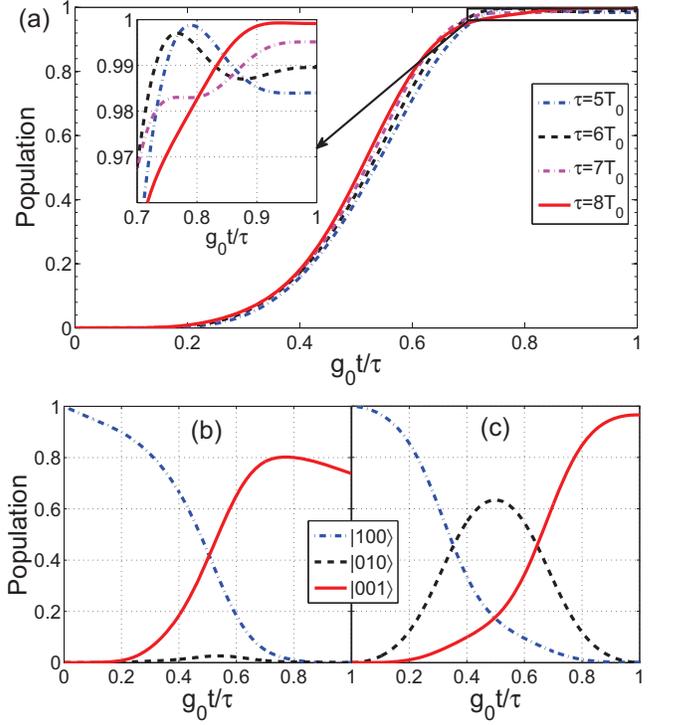}
\caption{(a) Fidelity dynamics of the adiabatic QSC for different time $\tau=nT_0 (n=5,6,7,8)$   without decoherence. Population dynamics of the QSC process for (b) adiabatic evolution $\tau=8T_0$  and (c) superadiabatic evolution $\tau=T_0$  cases  with decoherence.}
\end{center}
\end{figure}

\subsection{Numerical simulations}
For the time dependence coupling, we choose $g_1(t)=g_0\sin^2\left(\frac{\pi t}{2\tau}\right)$, $g_2(t)=g_0\cos^2\left(\frac{\pi t}{2\tau}\right)$. In order to obtain such coupling in opto-electro-mechanical systems, modulating the external driving field is a feasible way, as $g_i \propto \sqrt{n}$ with $n$ being the photon number inside a cavity. In the strongly-driven condition, external driving field excites large number of photons, $10^{6} \sim 10^{8}$, in the cavity \cite{photons}.  Therefore, the amplitude and phase of the couplings can be adjusted in a broad range \cite{tunable coupling}.  Note that during the superadiabatic correction process, one should guarantee that each corrected couplings $\tilde{g}_{1}(t), \tilde{g}_{2}(t)$ cannot exceed its original couplings' peak amplitude $g_0$, i.e., we need to ensure that $\max\left[\tilde{g}_{1}(t),\tilde{g}_{2}(t)\right]\leq g_0$. This constraint implies that we can only speed up the process with an minimal time  $T_0=3.24/g_0$, through our numerical verification. As shown in Fig. 2(a),   the final fidelity approximately reaches 98.4\%  and 99.9\%  for $\tau=5T_0$ and $\tau=8T_0$ without decoherence, respectively. Therefore, we set $\tau=8T_0$ as the adiabatic case for our reference.

We now compare the performance of the adiabatic and superadiabatic QSC under dissipation. The performance of the QSC is evaluated by considering the influence of dissipation using the Markovian master equation
\begin{equation}
\dot{\rho}=-i[H(t),\rho]+\frac{\gamma_1}{2}L(a_1)
+\frac{\gamma_2}{2}L(a_2) +\frac{\kappa}{2}L(b_m),
\end{equation}
where $\rho$ is the density matrix of the considered system. $L(A)$  is the Lindblad superoperator $L(A)=2A\rho A^\dagger-A^\dagger\rho A-\rho A^\dagger A$ with $A\in\{a_1, a_2, b_m\}$; $\gamma_1$ and $\gamma_2$ are the decay rate of the optical cavity $a_1$ and microwave cavity $a_2$ due to the loss of photons inside cavity; $\kappa$ is the decay rates of the mechanical oscillator $b_m$. Here, we take the decoherence induced by the mechanics mode for having thermal excitation at low temperatures. Here, we choose $g_0=2\pi\times 5$ MHz and the decay rates $\kappa=\gamma_2=g_0/1000$, $\gamma_1= g_0/50$ has already been demonstrated experimentally \cite{photons}, where a fidelity of  can be obtained. As shown in Fig. 2(b), when $\tau=8T_0$, we can only obtain a fidelity of 73.80\% for the QSC. Correspondingly, when $\tau=T_0$, one finds that the conversion fidelity reaches 96.53\% as shown in Fig. 2(c).

\subsection{Suppression of the intermediate state population}

In most dissipation system, operation time and decoherence are two major factor influencing the final fidelity. There is a trade-off between operation time and decoherence \cite{wang2}. When the operation time is long enough to satisfy the adiabatic condition well, high fidelity can be obtained, while dissipation will destroy it due to long time integral. When the operation time is too short, non-adiabatic leakage may lead to  poor performance during the conversion procedure. Through the superadiabatic correction, there is no need to worry about this trade-off, for we release the adiabatic condition. The decoherence property of the system becomes our major concern.

However, as shown in Fig. 2(c), the population of the intermediate mechanical mode is pretty large, which is what we should try to avoid. The intermediate state may decay to the ground state, and thus reduce the conversion fidelity. Therefore, if the decay of the intermediate mode is large, one of the main issue of QSC mediated by a quantum bus is to find ways of reducing the population of intermediate state. The population of the intermediate level is determined by $|\langle\psi(t)|b_m\rangle|^2=\sin^2\mu(t)$, i.e., we may try to reduce $\mu(t)$ in order to suppress the population. Therefore, we generalize Eq. (\ref{15}) to
  \begin{align}
    \mu^{'}(t)&=\arctan{\left[\frac{\dot\theta(t)}{f(t)g(t)}\right]}, \notag\\
    g_x(t)&=\dot\mu^{'}(t),  \notag\\
     g_z(t)&=-g(t)+\frac{\dot\theta(t)}{\tan\mu^{'}(t)},
  \end{align}
by introducing an auxiliary function $f(t)=1+A\sin^4\left(\pi t/\tau\right)$, with the squeeze parameter $A>0$  been optimized for each operation time. Note that the pulse shape is different from that of in Ref. \cite{dress}, and thus the auxiliary function is different.

\begin{figure}[tbp]
\begin{center}
  \includegraphics[width=8cm]{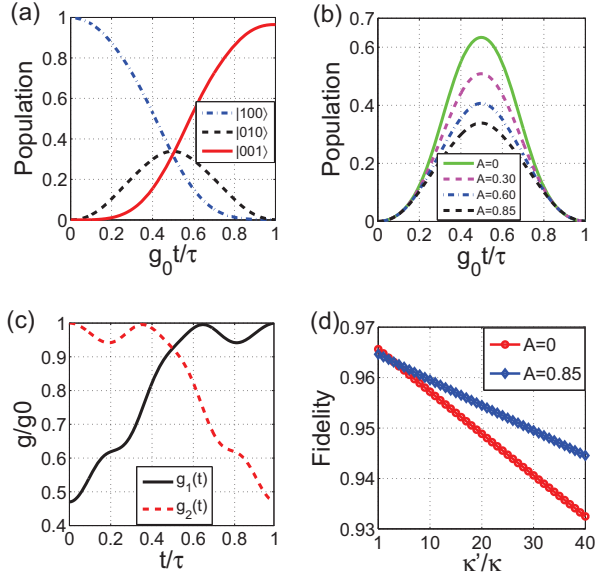}
\caption{(a) Population dynamics of the superadiabatic case with the conversion time to be $\tau=T_0$ and the squeeze parameter $A=0.85$.    (b) Population dynamics of the intermediate state for different $A$. (c) Pulse shape when $A=0.85$. (d) Final state fidelity with different decay rates of the intermediate mechanical mode with and without suppression of its population.}
\end{center}
\end{figure}

To illustrate the suppression of the intermediate state population, an auxiliary function $f(t)$ with $A=0.85$ for operation time $\tau=T_0$ is chosen as an example, as shown in Fig. 3(a),  a fidelity of 96.45\% can now be obtained and the maximum population of the intermediate state drops from 0.63 ($A=0$), as shown in Fig. 2(c), to 0.34 ($A=0.85$). While the largest $A$ we can get here is 0.85, for we need to guarantee that the peak value of $\tilde{g}_{1}(t)$ and $\tilde{g}_{2}(t)$ is no larger than the peak value of original strength $g_{1}(t)$ and $g_{2}(t)$, respectively. We also choose a set of $A$ range from 0.85 to 0, the population of the intermediate state witness a significant decrease as we increase $A$, as shown in Fig. 3(b), which shows our way of reducing the population of the mechanical mode is quite effective. The pulse shapes for $A=0.85$ is plotted in Fig. 3(c). Under this suppression, we note that the fidelity is deceased instead of increased. This is because  the decay of the optical cavity is the main decoherence source in our system, the suppression requires further modification of the pulse shape, which deviates from the optimal one and thus results in slight decrease of the final state fidelity. If the decay rate of the intermediate state is larger, the suppression of the immediate population will be more important, and the suppression will lead to the increase of the final fidelity, as shown in Fig. 3(d).

\begin{figure}[tbp]
\begin{center}
  \includegraphics[width=8cm]{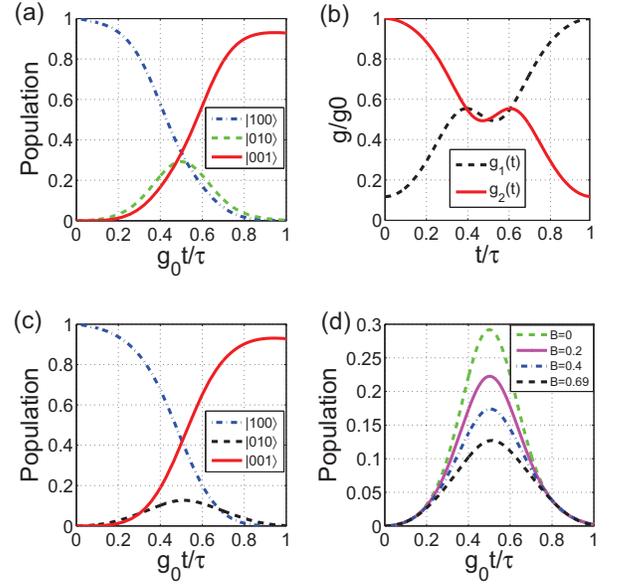}
  \caption{(a) Population dynamics of the superadiabatic process when $\tau=2T_0$ without suppression ($A=0$), and the pulse shape is plotted in (b). (c) Population dynamics of (a) with $A=0.69$. (d) Population dynamics of the intermediate mode for different $A$.}
\end{center}
\end{figure}

We note that when the operation time is longer, the population of the immediate mechanical mode can also be suppressed. Therefore, we further explore the conversion  fidelity for a slower process. For $\tau=2T_0$,  the fidelity is 92.86\% in Fig. 4(a) and the corresponding superadiabatic pulse shapes have been plotted in  Fig. 4(b). Meanwhile, by introducing another auxiliary function $f'=1+B\sin^4\left(\pi t/\tau\right)$ with $B=0.69$, the fidelity we can obtain is 92.81\%, as shown in Fig. 4(c). It is obvious that the fidelity is also slightly decreased as we expained in the above. Meanwhile, the fidelity is smaller that the case of $\tau=T_0$. This is quite natural as the decay rate of the mechanical mode is small, and thus the operation time here is the most important decoherence source. The largest $B$ we can choose for $\tau=2T_0$ is 0.69, since we still need to guarantee that peak amplitude of $\tilde{g}_{0}$ is no larger than the peak amplitude of $g_{0}$ . Hence, we also choose a set of A range from 0.69 to 0, the maximum population of the intermediate state also witness a significant decrease, as shown in Fig. 4(d).

\section{Conclusion}

In conclusion, we have proposed a scheme to realize the superadiabatic QSC process in opto-electro-mechanical system, which can significantly speed up the adiabatic procedure by using dressed state. Our scheme possess the following remarkable advantages. Firstly, there is no direct coupling between the target and initial modes in the Hamiltonian, and thus is feasible experimentally. Secondly, during the whole evolution, the adiabatic condition is released, and thus fast and high fidelity can still be achieved compared to the conventional ones. Therefore, our protocol presents an important step towards practical state conversion between optical and microwave photons, and thus may find many applications in hybrid quantum information processing.

\acknowledgements

This work was supported in part by the NFRPC (No. 2013CB921804), the NKRDPC (No. 2016YFA0301803), and the NSF of Jiangsu province (No. BK20140588).

\end{document}